\documentclass[aps,prd,twocolumn,nofootinbib,superscriptaddress]{revtex4}
\usepackage{amsmath} 
\usepackage{verbatim} 
\usepackage{mathrsfs}
\usepackage{amsfonts}
\usepackage{amssymb}
\usepackage{graphicx}
\usepackage{color}
\usepackage{hyperref}
\usepackage{soul}
\usepackage[utf8]{inputenc}
% ...
\inputencoding{latin1}
% ...
\inputencoding{utf8}

\newcommand{\nk}{\textbf{k}}

\newcommand{\x}{\textbf{x}}

\newcommand{\bra}{\langle}
\newcommand{\ket}{\rangle}

\newcommand{\nn}{\nonumber \\}

\begin{document}

\title{Discussions about the landscape of possibilities for treatments of cosmic inflation involving continuous spontaneous localization models}

\author{Gabriel R. Bengochea}
\email{gabriel@iafe.uba.ar}
\affiliation{Instituto de Astronom\'\i
	a y F\'\i sica del Espacio (IAFE), CONICET--Universidad de Buenos Aires, 1428 Buenos Aires, Argentina}

\author{Gabriel Le\'{o}n}
\email{gleon@fcaglp.unlp.edu.ar }
\affiliation{Grupo de Astrof\'{\i}sica, Relatividad y Cosmolog\'{\i}a, Facultad
	de Ciencias Astron\'{o}micas y Geof\'{\i}sicas, Universidad Nacional de La
	Plata, Paseo del Bosque S/N 1900 La Plata, Buenos Aires, Argentina}
\affiliation{CONICET, Godoy Cruz 2290, 1425 Ciudad Aut\'onoma de Buenos Aires, Argentina. }

\author{Philip Pearle}
\email{ppearle@hamilton.edu}
\affiliation{Emeritus, Department of Physics, Hamilton College, Clinton, NY 13323, USA}

\author{Daniel Sudarsky}
\email{sudarsky@nucleares.unam.mx}
\affiliation{Instituto de Ciencias Nucleares, Universidad
	Nacional Aut\'{o}noma de M\'{e}xico, A.P. 70-543, M\'{e}xico
	D.F. 04510, M\'{e}xico.}
\begin{abstract}
	
In this work we consider a wide variety of alternatives opened when applying the continuous spontaneous localization (CSL) dynamical collapse
theory to the inflationary era. The definitive resolution of many of the issues discussed here will have to await, not only for a general relativistic CSL theory, but for a fully workable theory of quantum gravity.  Our concern here is to explore these issues, and to warn against premature conclusions. This exploration includes: two different approaches to deal with quantum field theory and gravitation, the identification of the collapse-generating operator and the general nature and values of the parameters of the CSL theory. All the choices connected with these issues have the potential to dramatically alter the conclusions one can draw. We also argue that the incompatibilities found in a recent paper, between the CSL parameter values and the CMB observational data, are associated with specific choices made for the extrapolation to the cosmological context of the CSL theory (as it is known to work in non-relativistic laboratory situations) which do not represent the most natural ones.
\end{abstract}

\maketitle

\section{Introduction}

With the specific interest in explaining the emergence of the seeds of the observed inhomogeneous cosmic structure, generated by the so-called quantum fluctuations, the question concerning the applicability of spontaneous collapse theories (i.e. modifications of quantum mechanics designed to resolve the so called ``measurement problem'')\footnote{See for instance \cite{Maudlin95} for a classification of the possible alternatives to do so, and where the author stated the measurement problem in a formal and general way.} to the inflationary account, is a natural one. This is because, on the one hand, in the cosmological context the measurement problem becomes exacerbated.\footnote{J. Bell's work \cite{Bell81} is one of the first references where the fact that the quantum measurement problem worsens in the cosmological case was noted.} On the other hand, without some modification, the usual account of the emergence of those cosmic seeds suffers from serious shortcomings \cite{Shortcomings}.

Various applications involving different versions of the continuous spontaneous localization (CSL) collapse theory \cite{Pearle1989,GPR,bassi} to the problem were attempted early on \cite{jmartin,Das2013,pedrocsl,LB15}. In particular, the last three approaches were able to recover the correct CMB power spectrum by making various assumptions regarding the dependence of the model parameters on time or on the Fourier mode wavelengths (the latter could be attributed also to a choice of a peculiar collapse operator). However, a recent paper \cite{jmartinPRL} concludes that the direct application of the CSL theory to the cosmological context, with a particular extrapolation from its non-relativistic form, is ruled out by observational data (except for very specific and fine tuned choices), ``casting a shadow" on models based on CSL theory.

In this work, we will explore some of the vast theoretical landscape available for the extrapolation of the standard CSL theory (as constructed to deal with non-relativistic many particle quantum mechanics), into the realms of relativistic quantum field theory in curved spacetimes, as would be required for applications in the inflationary context. We will also place into a broader perspective both the approach as well as some of the conclusions of \cite{jmartinPRL}, arguing that the analysis carried out does not represent either unique or even natural choices for the generalizations required. Thus, such conclusions should not be considered as broadly reliable, as they rely on a large set of both explicit and implicit assumptions, which might be considered as questionable for several reasons that will be discussed below.

Incorporation of collapse theories into regimes involving both general relativity (GR) and quantum field theory (QFT) are actually densely fraught with difficult issues, and therefore any conclusion on the direct application of the non-relativistic CSL theory to the case of inflation must be considered as merely exploratory at best. However, without some kind of modification to the standard Quantum Theory, such as that offered by CSL, the account of the emergence of primordial inhomogeneities is not really satisfactory, a fact acknowledged in \cite{jmartinPRL}. This issue is associated with the so called ``quantum-to-classical transition'' which, as described by the authors in \cite{jmartinPRL}, {\it ``becomes even more acute in cosmology than in the laboratory, due to the difficulties in introducing an `observer' as in the standard Copenhagen interpretation''}.

Throughout this article we will focus on three main aspects: first, in section \ref{secdos} we will compare two different frameworks in which the treatment of the interface between gravitation and quantum field theory is carried out. After that, in section \ref{sectres} the choice of the collapse-generating operator will be discussed, and third, in section \ref{seccuatro}, we will analyze the nature and values of the fundamental collapse parameters of CSL type models. Finally, in section \ref{conclusions} we present our conclusions. Regarding conventions and notation, we will use units where $c=1=\hbar$.

\section{Gravitation and Quantum Field Theory}
\label{secdos}

Since collapse theories have been constructed in terms of modifications of the Schr\"odinger equation, which basically characterizes the time evolution of the system quantum state, one needs to work with such theories in a framework where time and space can be taken as sufficiently well defined (even if ultimately one holds the view that spacetime is something emergent from a deeper quantum gravity framework). There are two general settings for putting together quantum field theory and gravitation:
\begin{itemize}
	\item[(i)] This approach involves choosing a background classical spacetime and even a classical background matter field(s), and then quantizing the perturbations of both the metric and the quantum field(s).
	
	This view suffers from a rather serious conceptual shortcoming: QFT is fundamentally grounded in a construction of field operators with   commutation relations that follow the causal structure of the underlying spacetime. In other words, causality is embedded into the framework by the requirement that quantum field operators, at space like separated events, must necessarily commute both in flat or curved spacetimes (see for instance \cite{Causal1, Causal2}). The problem with this approach is that, as the metric itself (or part thereof) is subjected to quantization, there is simply no way, even in principle, in which those basic tenets could be rigorously applied. So, one is not really respecting the basic premises of QFT in curved spacetime.
	
	The second issue is that the separation of background and perturbations is, in a certain sense, extremely arbitrary. This is known as the gauge problem in perturbation theory, as applied to relativistic theories. The work of Bardeen \cite{bardeen} is usually taken as resolving the issue by constructing gauge invariant quantities. The problem is, however, that such quantities are only gauge invariant if one restricts oneself to infinitesimal gauge transformations, starting from an in-principle arbitrary initial choice\footnote{Note that the issue is in general much more serious than the simple consideration of what happens if one wants to go beyond the lowest order in the analysis} \cite{Stewart74, Adolfo11}. In other words, as the quantities involved are not truly gauge invariant under general finite transformations, the trouble of fixing the original gauge remains, in general, an open one. At the conceptual level, the situation becomes rather problematic  when the quantized quantities are a combination of metric and matter perturbations. This is because even if one focuses on infinitesimal gauge transformations (which leave the so-called Bardeen variables unchanged), the fact that those variables are composed from metric and matter field perturbations implies that, when considering a gauge transformation, what is described quantum mechanically and what is not will have to be regarded as changing.
	Therefore, there could be some gauge in which one of the quantities treated in a quantum language refers just to matter fields, whereas in another gauge the same quantity might include also a part of the metric. Thus, while working with such ``gauge invariant quantities'' might be convenient at the practical level, questions about their physical meaning (or in other words their ontological status \cite{Bell81}), become even more obscure. In this sense, working with a fixed gauge might be less ``elegant" but it seems conceptually clearer.

With all of this in mind, one is compelled to wonder about the foundations of this approach and recognize that they are, at least, seriously questionable. However, in the cosmological setting these issues are usually ignored (in part because there are natural choices where the perturbations are always very small, $\sim 10^{-5}$). This approach is widely employed in the literature by many authors, including in \cite{jmartinPRL}.

	\item[(ii)] The second path is to take the semiclassical gravity framework\footnote{Note that since we are dealing with the description of the inflationary epoch, the energy scales involved allow one to suppose valid the consideration of the metric as classical and well described by semiclassical gravity equations.}.
	
	In this case, the full metric is characterized in a completely classical language, while all matter fields, not just perturbations, are described by using QFT in curved spacetime. Semiclassical gravity is characterized by Einstein semiclassical equations
	\begin{equation}\label{SCGEE}
	G_{ab} = 8 \pi G \bra \hat T_{ab} \ket,
	\end{equation}	
	where $\hat T_{ab}$ is the energy-momentum tensor (operator) associated to the matter quantum fields, $\bra \hat T_{ab} \ket $ its renormalized expectation value in a suitable quantum state of the quantum fields and $G_{ab} \equiv R_{ab} -  g_{ab} R/2$ is the Einstein tensor.
	
This approach has been the subject of criticisms and even arguments that it is nonviable \cite{unviable-semicalsssical1,unviable-semicalsssical6}. Those arguments have been refuted \cite{unviable-semicalsssical2,unviable-semicalsssical5,semicalsssical-and-collpase2} (although those refutations have sometimes received less attention). In particular, some of us have been involved in the development of a formalism allowing one to rigorously incorporate spontaneous collapse models into semiclassical gravity; a program that, while not yet finished, has led to important advancements \cite{alberto,erandy,benito}. Nonetheless, we acknowledge that this second program can of course also be subjected to additional criticisms and scrutiny, where subtle questions might arise.

\end{itemize}

For a long time, naive expectations have been that approaches (i) and (ii) would generally yield similar results. To shed light on the issue, some of us decided to carry out various
analyses within framework (i). We found that, by including collapses, the results are different from the standard case, but also that, under certain particular conditions, similar results can be obtained\footnote{However, as discussed in e.g. \cite{Shortcomings}, the standard framework without collapses cannot account for the emergence of any primordial inhomogeneities or anisotropies, simply because the theory does not include any physical mechanism to break the corresponding symmetries present in the initial vacuum state. On the other hand, we also compare the results obtained between approaches (i) and (ii).}. For instance, in \cite{Das2013,LB15,Bouncing16} it was shown that very particular CSL types of models, based on approach (i), could be consistent with both the CMB observations and constraints coming from non-relativistic laboratory situations\footnote{After the introduction of suitable assumptions regarding the scale dependence of the model parameters, which in turn, might be viewed as tied to a more fundamental dependence on curvature as will be discussed latter.}. Therefore, in a certain sense, even omitting the serious problems mentioned in (i), the results in \cite{jmartinPRL} should be interpreted as merely ``casting a shadow" on another particular type of CSL model, which is different from the ones considered in \cite{Das2013,LB15,Bouncing16}. In other words, even in approach (i), which we deem as problematic at the conceptual and theoretical level, one can still find CSL inspired models that are not in conflict with the empirical data. This disagrees with the overreaching conclusion by the authors in \cite{jmartinPRL}.

A very interesting result that distinguishes the two approaches is that the predictions regarding primordial gravity waves, manifested in the B-modes polarization of the CMB, are radically different in the two approaches. In particular, according to (i), most inflationary models \cite{guzzetti2016} predict that the current searches should have already detected such modes, in conflict with their, up to this date, experimental absence \cite{planck2018inflation}. However, the generic prediction emerging from approach (ii) indicates that the overall expected spectral amplitude of these so-called tensor modes will appear with a substantial suppression (to the point that it is natural to expect their non-detection at current attainable levels). Moreover, the primordial B-modes might be expected to have a radically different spectral form from those presently experimentally accessible, so that any hope to detect them would have to focus on the very low angular multipoles, in comparison with the rather scale-free character of the primordial spectrum for the scalar modes observed in the CMB \cite{nobmodesshort, nobmodesbig}.

So far, we have given two problems faced by approach (i) not experienced by approach (ii): that the metric perturbations are not subject to collapse yet, for consistency, they should be treated in combination with the matter fields which are subject to collapse, and prediction of the undetected B-modes polarization.

We now turn to consider differences between approaches (i) and (ii) concerning the emergence of the primordial inhomogeneities/anisotropies.

In case (i) one must make extra assumptions to justify the required replacement of the quantum fields, associated to the metric perturbations, by their classical (and stochastic) counterparts. For example, one might consider that it could be justified to identify the expectation value of the quantum field associated to a scalar metric perturbation, say $\bra \hat \zeta \ket$, with a corresponding classical field $\zeta$ (as inspired by Ehrenfest's theorem), the latter representing the curvature perturbation in a particular gauge. Of course, the issue is what exactly is the nature of such justification, and under what conditions might it be said to hold in this context. On the other hand, in approach (ii), no additional assumption of this kind seems to be required at all, since the metric perturbation is always a classical field.

Another distinction is that, according to approach (i), the authors of \cite{jmartinPRL} consider that a certain mode exhibits a presence substantial enough to count as a seed of structure only if the collapse brings about a wave function characterized by the expectation value of the relevant quantity that is substantially greater than its corresponding width; i.e., that it has become something that might be \emph{``effectively described in classical terms"}.

In the context of (ii), the emergence of well-defined inhomogeneities in the metric does not require that the state of the inflaton be sharply peaked. It is enough that the non-homogeneous components develop a non-vanishing expectation value. The fact that the metric perturbations are controlled by the expectation value of the energy-momentum tensor via $\delta G_{ab} \propto \bra \delta \hat T_{ab} \ket$, implies that the value of the uncertainty of the latter does not seem to have any direct implication on the degree of inhomogeneity and anisotropy of the spacetime.

In order to illustrate this point, we consider the work \cite{multiples} by some of us. There, a toy model based on a set of multiple discrete collapses is analyzed (which is thus not an example of CSL), with the requirement that, after each collapse, the resulting state be characterized at the time by a coherent state. As is known, for a coherent state the field and momentum variables have the same uncertainties as the vacuum state but, unlike the latter, the corresponding expectation values generically do not vanish (the vacuum state is a particular coherent state where these expectation values do vanish, see e.g. \cite{walls}). Using approach (ii), and in particular, the perturbative version of Eq. \eqref{SCGEE}, one relates the classical metric perturbations to the expectation value of the perturbations of the energy-momentum tensor components. The results of \cite{multiples} showed that, due to the assumed finite number of collapses in that model, acceptable primordial metric perturbations are present, rendering the spacetime inhomogeneous and anisotropic, and in principle compatible with the known spectrum. In addition, we obtained a consistent shape of the primordial spectrum. In particular, the final result is characterized by coherent states, hence all quantum uncertainties of the field variables are the same as the vacuum state, but the spacetime and each post-collapse state lost its original symmetries (i.e. the homogeneity and isotropy), which is an empirically acceptable characterization of the situation. Various works \cite{PSS,adolfo2008,multiples} following this approach (relating metric perturbations to the expectation value of the energy-momentum tensor components) show that, generically, the collapses in the matter sector lead to the emergence of primordial metric perturbations, thus rendering the spacetime inhomogeneous and anisotropic. Given such conditions the subsequent evolution of matter might be expected to follow the standard expectations.

Thus, the criteria used in \cite{jmartinPRL} for the primordial inhomogeneities to be generated (see condition given in Eq. (7) of that Ref.) are not the appropriate ones for approach (ii). Importantly, we will argue that this has very relevant implications regarding what the values of the CSL parameters should be, and whether or not they are compatible with CMB observations.

In this latter regard, one must keep in mind that, in both approaches, what one observes in the CMB is not a direct manifestation of the inflaton field inhomogeneities, as characterized by their quantum state of the corresponding modes $\nk$. Rather, it is the result of their further evolution and transmutation in fluctuations of the matter distribution (ordinary matter such as quarks, leptons and photons, as well as the still mysterious dark matter component).

Let us be more specific and introduce the function $\Delta T(\hat n)$, representing the CMB temperature fluctuations relative to a background temperature $T_0 \simeq 2.7$ K, where $\hat n$ denotes a direction in the sky. The expansion of that temperature map in spherical harmonics is
\begin{equation}\label{tempYlmexp}
\Theta (\hat n) \equiv \frac{\Delta T (\hat n)}{T_0} = \sum_{l,m} a_{l m} Y_{lm} (\hat n),
\end{equation}
where $Y_{lm}$ are the standard spherical harmonics defined on a two-sphere. Thus,
\begin{equation}\label{alm0}
a_{lm} = \int d\Omega \: Y_{lm}^\star \Theta (\hat n).
\end{equation}

Then, one introduces the so called \textit{transfer function} $\Delta^T_l (k)$, which takes into account the evolution (from the end of the inflationary period until today) of the particular Fourier mode associated to the scalar metric perturbation $\zeta_\nk$ (in a particular gauge), and its effect on the temperature variation $\Delta T$. Consequently, the $a_{lm}$ coefficients are given by
\begin{equation}\label{alm1}
a_{lm} = 4 \pi (-1)^l \int \frac{d^3 k}{(2 \pi)^3} \Delta^T_l (k)  Y_{lm} (\hat k) \zeta_\nk.
\end{equation}
Thus, in Eq. \ref{alm1} one can clearly see the relation between the observational quantities $a_{lm}$ and the quantity denoted by $\zeta_\nk $.

The issue we must contend with is that such a relation might become problematic in treatments that contain simultaneously classical and quantum objects. In fact, within approach (i) this is a very serious problem, because $\zeta_\nk$ is a quantum mechanical operator (i.e. a Fourier transform of a quantum field operator). Therefore, Eq. \ref{alm1} would not make sense unless we take $a_{lm}$ to be a quantum operator as well.\footnote{Such a posture would also be problematic, unless one were willing to accept that we actually observe quantum operators. This is certainly not what orthodox quantum mechanics holds. What it says is that we measure eigenvalues of such operators, corresponding to the state into which the system collapsed, as a result of our observation. And as it has been widely discussed elsewhere, such a posture is simply not tenable in the situation at hand.} The simplest and most natural possibility would be to replace $\zeta_\nk$ with its corresponding expectation value $\bra  \hat \zeta_\nk \ket$ in a suitable quantum state. But, unfortunately, for all states contemplated in the standard approach (particularly the Bunch-Davies or similar ``adiabatic" vacuum states) such quantity vanishes.

What is often done, is to claim that the quantity in question should be identified with the corresponding widths or quantum uncertainties $\bra \hat \zeta_\nk^2 \ket^{1/2}$. However, a closer analysis clearly indicates that $\bra \hat \zeta_\nk^2 \ket^{1/2}$ cannot be directly taken as indicating anything regarding possible \emph{fluctuations} in the observable quantities. A simple way to see this, is to note that taking such identification seriously, one would be forced, among other things, to conclude that there could be absolutely no difference between the various $a_{lm}$ for a fixed value of $l$ but different values of $m$'s, since all terms appearing in the RHS of Eq. \eqref{alm1} are rotationally invariant except for $Y_{lm}$. In fact, the latter observation alone implies that $a_{lm}$ vanishes, except for $(l, m) =(0, 0)$, if the quantum uncertainties $\bra  \hat \zeta_k ^2 \ket$ are isotropic.

One possibility is to make the identification $\zeta_\nk = x_\nk \bra \hat \zeta_\nk^2 \ket^{1/2}$, with $x_\nk$ a Gaussian random variable (whose distribution is centered at zero and width unity). However, the fundamental origin of the stochastic nature of $\zeta_\nk$ would then be unclear, unless one considers a collapse type of model. This is because the evolution equations in standard quantum mechanics are fully deterministic, and the only element that might induce a stochastic aspect to the system is given only when there is a ``collapse of the wave function''. Therefore, whether or not $\zeta_\nk$ vanishes, and if not, whether it actually displays a stochastic behavior, depends entirely on whether or not the collapse is triggered, and has nothing to do with the existence of quantum fluctuations/uncertainties.

The approach adopted in \cite{jmartinPRL} attempts to address this issue, by retaining approach (i), but it modifies the quantum dynamics with the incorporation of CSL applied  directly to a quantum field that includes the metric perturbation. In this way, the quantity $\zeta_\nk$ in Eq. \eqref{alm1} is identified with $\bra  \hat \zeta_\nk \ket $, evaluated in the quantum state that results from that modified evolution. According to \cite{jmartinPRL}, said  identification would only make sense  if the results from the CSL dynamics are such that the quantum uncertainty associated with the operator $\zeta_\nk$, is much smaller than its expectation value (i.e. the state can be thought of having a sufficiently sharply defined value of the quantity in question). And therein lies what, according to the analysis of Ref. \cite{jmartinPRL}, CSL fails to do. Their argument against CSL is that, using the values of the parameters that characterize its low energy, non relativistic particle regime, it fails to achieve a sufficient localization of the relevant wave functions in the inflationary context.

We note that, when following approach (ii), the issue looks very different. According to semiclassical gravity, the full metric (perturbations included) is always treated in classical terms, and it is sourced by the expectation value of the energy-momentum tensor. Thus $\zeta_\nk$ is never a quantum mechanical object because it is just the  Fourier transform of a metric perturbation. Thus, the necessity of resolving the difficulties that might be faced in (i) are never encountered. At worst, one might question under what conditions it is reasonable to trust the analysis based on semiclassical gravity. We will delve more deeply into this question a bit later in the manuscript but, for the moment, we will continue analyzing the considerations underlying \cite{jmartinPRL}.

A further problematic issue with the analysis of \cite{jmartinPRL}, as well as most analyses of the subject based upon (i), is that they usually consider such questions on a mode by mode basis. However, such modes, characterized by wave numbers $\nk$ and having a spatial behavior given by the standard factors $e^{i\nk \cdot  \x}$, are not spatially localized. Their standard argument, suggesting that one should view each mode as somehow representing a fluctuation of size $k^{-1}$ [multiplied in this case by the cosmological scale factor $a(t)$], is based on a simplistic argument related to Heisenberg's uncertainty principle.

Let us consider the simple case of one non-relativistic particle in one dimension in quantum mechanics. The rigorous statement is that $\Delta X  \Delta  P  \geq 1/2$ (where $\Delta O \equiv [\bra \hat O^2\ket - \bra \hat O \ket^2]^{1/2}$ is the standard deviation associated to a Hermitian operator $\hat O $) for all states. So, if we try to construct a system at rest which is rather well localized up to a particular tolerance length $L$, i.e. $\Delta X <  L$, we have $\Delta P > L^{-1}/2 $. Also, if the wave packet in momentum space is centered at $P=0$, then it must have ``substantial  support'' up to values of $P$ including $|P| > L^{-1}/4$. However, by looking at this wave packet in a boosted frame, where its momentum is centered at $P_0 \gg 0$, it is clear that such a value of $P_0$ has nothing to do with how well localized in $X$ the system might be at any given time. In other words, the notion that a mode with wave number $\nk$ should be considered as spatially localized within a region of size $k^{-1}$ is simply not correct. In fact it is quite clear that any mode with spatial dependence given by $e^{i\nk \cdot  \x}$, remains absolutely unmodified by any displacement $\Delta \x $, such that $\nk \cdot  \Delta \x =  2\pi N$ with $N$ any natural number. So much for its localization!

Modes $u_L$ that are localized in space within regions of size $L$ should be constructed from a combination of modes $\nk$, involving a wide enough range $\Delta  k$, as prescribed by the well known features of the Fourier transform. Moreover, these modes should be excited in a certain correlated way and, in particular, their relative phases should bear some level of regularity. Now, apart from the well known issues that appear in general context in Minkowski spacetime, as Haag's theorem and its implications about the non-existence of the interaction picture, the general characterization of the localization field excitations in view of the Reeh–Schlieder theorem \cite{reeh61}, and the related difficulties with candidates for position operators including the Newton-Wigner proposal\footnote{Needless is to say that none of those issues detracts in any way from the stupendous success of QFT, and in particular the Standard Model, as applied to the practice at the calculational and experimental levels of particle physics}, serious further difficulties arise in QFT in curved spacetime.

For example, when considering the general notion of an excitation's localization in Minkowski spacetime, one might hope, for instance, to take the point of view that the field excitation is localized in the spacetime region where the expectation value of the energy-momentum tensor (EMT) is non-vanishing (or where that quantity differs from zero by a sufficient amount). But it is far from evident how exactly such a requirement should be expressed with a particular level of localization, in field space, of the quantum state in terms of each of the modes involved, given the fact that for general states the various modes will be highly entangled. Moreover, in passing to curved spacetimes, there are various new aspects one must contend with: a) The EMT is not a well defined operator, and the evaluation of its expectation value at a point requires renormalization, a procedure  known to involve certain inevitable ambiguities \cite{Wald94}. b) A reasonable aspect seems to expect that the characterization of the localization of an excitation associated with a certain quantum state, to include also the corresponding uncertainties, besides the expectation value of the EMT. We note that while there is a lot known about renormalization of the expectation value of the  EMT, much less is know about the renormalization of  its  uncertainties. c) One must generate some well defined scheme to associate some overall level of non-localization to the quantum state of the field in terms of the values of those quantities over extended spatial regions. This aspect will likely require some general scheme to specify some adequate foliations of spacetime.

We have presented some concerns associated with the argument of \cite{jmartinPRL} (and will present others) in order to make the case that the conclusions arrived at by these authors can have no definitive status. Indeed, there are quite a few unsatisfactorily resolved issues regarding combining gravity and QFT in any present-day approach. That of (ii) is no exception, requiring the following critical remarks.

In what follows below, we will focus on the issue of conditions where one might or might not trust semi-classical gravity. A point of view that is quite generally accepted concerning semiclassical gravity is that it is considered reliable only when the quantum state has a relatively well defined energy-momentum tensor, and not when one faces, for instance, a state representing a large amount of matter in a superposition of distant but sharply localized states. But, proper inclusion of collapse dynamics should rapidly remove such a superposition. However, there is still the complex question of reliability of the semi-classical theory in contexts in which the state is represented by a single localized but rather wide wave function.

We start by noting that one cannot take the position that the theory can only be trusted when the expectation values of the anisotropic and inhomogeneous parts of the energy-momentum tensor are much greater than the corresponding widths, analogous to what is done in \cite{jmartinPRL} with their choice of energy density operator. If we adopted this view, then we would find ourselves in a situation in which we could not trust the theory even in the case where the $k\not =0$ modes are in the Bunch-Davies vacuum (i.e. before any collapse took place), simply because in those conditions the expectation value associated with such modes vanishes, while the corresponding uncertainties do not. Therefore, it appears undesirable to require as a condition on viability of the semiclassical treatment that, on a mode by mode basis, the expectation value of the energy density be greater than its corresponding quantum uncertainties, in analogy to what is done [while working within the approach (i)] by \cite{jmartinPRL}. Instead, one might require that the expectation value of the energy-momentum tensor in spacetime be greater than the corresponding quantum uncertainty. But, then,  it is not clear how precisely that requirement should be imposed.

We certainly cannot do this point by point since, due to the distributional nature of the quantum field, the quantum uncertainties in the expectation value of the energy-momentum tensor at one point are formally divergent.\footnote{In fact the expectation value of energy-momentum tensor is itself formally divergent. But in this case, and in contrast with the quantum uncertainties, the  quantity in question is at least renormalizable provided we are considering a Hadamard state.} Rather, it seems reasonable that the required condition be that the expectation value of the energy-momentum tensor, smeared over a certain sized regions and with certain smooth functions of compact support, be greater than the corresponding uncertainties.  On first  sight one  might mistakenly  think  that even this condition  would fail to hold  in our case, because in the Bunch-Davies  (or adiabatic) vacuum, the renormalized energy-momentum vanishes. However one must recall that during inflation (even before the effects of CSL are taken into account) not all modes are in the vacuum. The zero mode of the field (which according to approach (ii) is treated quantum mechanically) is in a highly excited state, thus, providing the overwhelmingly dominant  contribution to the energy-momentum tensor. In comparison to that, it might well be that with a suitable renormalization procedure (yet to be  developed) the contributions to the above discussed local measure of the quantum uncertainties from all the $k\not =0 $ modes might turn out to be negligible, both for the initial state as well as for any of the states that occur after including the CSL effects. After all, it is the zero mode contribution that drives the most salient gravitational feature of the   situation at hand, namely the de Sitter-like accelerated expansion. In that case, attempts to extrapolate into approach (ii) the arguments employed in \cite{jmartinPRL}, regarding the level of uncertainties associated with the individual $k\neq 0$ modes for the state resulting from the  CSL dynamics, would be aimless.

\section{Collapse-Generating Operator}
\label{sectres}

The next issue that must be confronted in this program is the identification of the collapse-generating operator(s). That is, the CSL term added to the Schr\"odinger equation contains, besides its randomly fluctuating classical components, a set of completely commuting operators. When the initial state vector is written as a superposition of the joint eigenstates of these operators, the collapse proceeds (if one neglects the Hamiltonian part of the evolution) by evolving the state vector toward one of these eigenstates, with probabilities in agreement with the Born rule. In the specialization of the CSL formalism to the non-relativistic CSL model, the operators are the mass-density at each  point of space.

As is well known, the energy density is not a relativistic invariant concept, and the energy density contrast is even worse in this regard. But, this is the entity used in \cite{jmartinPRL} for the collapse-generating operator.

The energy density contrast is defined as $\delta \rho/\overline{\rho}$ where $\delta \rho$ and $\overline{\rho}$ correspond to the perturbed and background parts of the energy density. In \cite{jmartinPRL}, the collapse-generating operator is associated with the perturbed part of the energy density $\delta \rho_g$ and then related to the gauge invariant quantity $\zeta$ as $\delta \rho_g/\overline{\rho} = \epsilon_1  \zeta - \epsilon_1 (1 + 3 \epsilon_1 a^2 H^2 \delta^{-2}) \zeta'/(3 a H)$. Although $\delta \rho_g$ and $\zeta$ are gauge invariant quantities [in the sense described in (i)], they appear to be naturally identifiable directly with standard physical properties characterizing the actual degrees of freedom, only in \textit{different} gauges. Concretely, in the Newtonian gauge, $\delta \rho_g$ is associated with the 0-0 component of the inhomogeneous (perturbed) part of the energy-momentum tensor, while in the comoving gauge $\zeta$ represents the commonly called ``curvature perturbation'', i.e. the Ricci scalar associated to spatial 3D-hypersurfaces. This blatant inconsistency illustrates that even when working with \emph{gauge invariant quantities} as characterized in cosmological perturbation theory, there can be issues concerning the physical meaning of such quantities and the relations among them.

Moreover, the above choice is not necessarily obvious nor is it uniquely qualified. There are a large number of operators that \textit{are} relativistically invariant and do reduce, in the simple non-relativistic regimes encountered in laboratory situations, to the required ``mass density" or ``energy density''. Some simple examples we might consider are $T^a_{\ a}$ the trace of the energy-momentum tensor, or the scalars $(T^{ab} T_{ab} )^{1/2}$, $(T^{ab} T_{bc} T_a^{\ c})^{1/3}$, etc.  A point that needs to be made is that any of these choices of operators, which are constructed using the energy-momentum tensor, will require substantial amount of work to assure one is dealing with the appropriate renormalized versions of the expressions under consideration. Evidently, this is an issue that involves substantial complexities. Nonetheless, it is clear that there are in fact a large number of other options available.

One of the possible options that we find quite intriguing is to have the energy-momentum tensor itself act as the collapse operator. That is, instead of having the CSL model operate with a scalar function at each spacetime point, we would have a full tensor appearing in the CSL dynamics. Of course, that would imply that the stochastic scalar function
(also called the white noise field) occurring in traditional versions of the CSL theory would be replaced by a stochastic tensor field. The choice of the energy-momentum tensor acting as the collapse operator represents a departure from traditional CSL constructions (although discussed early on in \cite{GPR}) since one usually deals at each time with a set of mutually commuting operators. In the case when $T_{ab}$ is taken as the collapse operator, that part of the recipe would be violated because the different components of the energy-momentum operator do not commute among themselves. This brings about complications that are similar to those that often arise when using the CSL theory, since usually the Hamiltonian and the collapse generating operators do not commute, so collapse behavior is accompanied by other effects. However, even in this case, one can expect that in the non-relativistic situations the  dominant role among the set of collapse operators would be played by the energy density in the rest frame of the piece of matter that is under consideration.

To summarize, the choice of collapse-generating operator made in Ref. \cite{jmartinPRL} has consistency issues and is neither the only choice nor necessarily the best choice. Therefore, conclusions based upon it should be viewed somewhat skeptically.

\section{The Parameters}
\label{seccuatro}

The next question is how, in the context of the early universe, to treat the fundamental parameters of the non-relativistic CSL theory, namely the
collapse rate parameter $\lambda$ and the size of the localization region $r_c$. While in \cite{jmartinPRL} it is assumed they have the same values as the ones in the non-relativistic setting, here we will discuss and again emphasize that this choice is not only far from unique but also not at all evident.

Let us start by recalling the basic elements in the non-relativistic CSL theory and how its parameters are introduced. In the state vector evolution equation, the parameter $\lambda$ and the operator $\hat A({\bf x})$ appear. The parameter $\lambda$ is the collapse rate for a neutron in a spatially superposed state,  and
\begin{equation}\label{smeared}
\hat A({\bf x})=C\frac{m}{M_{N}}\int d{\bf z}e^{-[{\bf x}-\bf{z}]^{2}/4r_{c}^{2}}\hat N({\bf z}),
\end{equation}
where $\hat N({\bf z})=\hat \xi^{\dagger}({\bf z})\hat \xi({\bf z})$ is the particle number density operator, constructed from (non-relativistic) creation and annihilation operators.
On the other hand, $m$ is the mass of the corresponding particle species (with $M_N$ the mass of the neutron), where such mass-proportionality is suggested by empirical evidence \cite{Pearle1994}, and the smearing length $r_{c}$ is the second parameter of the theory. The collapse-generating operator $\hat A({\bf x})$ (toward one of whose eigenstates the collapsing state vector tends), may be thought of as a \textit{smeared mass density} operator at ${\bf x}$. The values $\lambda\approx 10^{-16}$ s$^{-1}$ and $r_{c}\approx 10^{-5}$ cm, proposed by Ghirardi, Rimini and Weber \cite{GRW} for their SL theory of instantaneous collapse, were tentatively adopted by Pearle in \cite{Pearle1989} for his CSL theory of continuous collapse, as providing sound behavior when applied to laboratory situations.

An argument for adopting those values is as follows. A basic test of a collapse theory is that a state vector describing a superposition of a ``pointer'' in two places (an approximation to the state of an apparatus for an experiment leading to two possible outcomes) should collapse toward one or the other pointer with a time scale that is shorter than human perception time, say $\approx 0.1$ s. The relatively slowest collapse of the smallest visible pointer is the most stringent test. When the size of the pointer is a small multiple of $r_{c}$ ($4\times r_{c}$ is the wavelength of blue light), the CSL prediction of this time is $T=1/\lambda N^{2}$, where $N$ is the number of particles in the pointer. With typical nuclear matter density a multiple of 1 g/cm$^{3}$, so $N\approx 10^{24}r_{c}^{3}\approx 10^{9}$, the GRW choice of parameters gives $T\approx 0.01$ s.
\textit{There is no reason why there should be any comparison between the values of these parameters, informed by the existence of  hadrons,  nuclei and  the  atomic  densities of solids, with parameters at an era when there were no nucleons,  no atoms  and  certainly no solids.}

Now we address the application to inflation of the present-day CSL localization structure, taken over wholesale in \cite{jmartinPRL}. In quantum field theory in curved spacetimes (QFTCS), the notion of particle disappears at the fundamental level \cite{Wald94}, leaving only the quantum fields themselves. Therefore, the notion of ``localization'', when considering a quantum field, is quite different than that pertaining to a particle. That is, the main quantum uncertainties characterizing the state of a particle are connected with its position in space (and its momentum), making the appearance of the fundamental length $r_c$ natural. The uncertainties characterizing the state of a quantum field are instead connected with the {\it ``value of the field''} at each point (and the conjugate momentum), so the corresponding natural parameter must have the same dimensions as the corresponding field, call it $\Delta$. Any relation that might exist between the parameter $r_c$, and $\Delta$ is then far from clear.

A similar problem arises when considering the collapse rate parameter $\lambda$, characterizing phenomena at laboratory scales. In relating the two regimes (laboratory experiments and the early inflationary era) it must be recognized that the universe passes through several important phase transitions, including nucleosynthesis, hadronization, the electro-weak transition, and more uncertain regimes such as reheating and the end of inflation. Thus, there are ample grounds to doubt any simple connection between the values of the parameters $r_c$ and $\lambda$, relevant for one regime, with the parameters characterizing the theory in a completely different one.

As an illustrative example of the implications of such phase transitions, consider what is known (which serves at an effective level to characterize the interaction between hadrons) as the strong nuclear force, which is now understood to be nothing but a small left over remnant of the $SU(3)$ color force between quarks, after these have combined into color singlets which constitute the commonly observable hadrons: protons, neutrons, hyperons, pions and so on. According to our current understanding, in the very early universe there were no hadrons at all. And during a relatively wide regime (in temperature scales) between the end of inflation and the hadronzation epoch, the related entities present in the universe were described by a quark-gluon plasma, characterized by completely different degrees of freedom and by a completely different dynamics than the corresponding ones prevailing after hadronization. Needless is to say that important parameters, like the proton, neutron and pion masses, as well as their effective coupling constants, bear a rather complex (and still not fully understood) relation with the relevant parameters of the theory in the pre-hadronization stage, such as quark masses and the color force coupling constant. The complex connection between the two regimes goes well beyond that relatively simple feature known as the running of that latter constant with energy, which is understood to account for the asymptotic freedom of the $SU(3)$ color theory.

Let us look at this issue in a little more detail. First of all, note that the role of the parameter $r_c$ is to smear the mass density operator over a spatial volume $r_c^3$. But then, again, smearing over a particular spatial volume involves notions that clearly depart from relativistic considerations. In general situations, one might identify a spatial volume only in the context of a specific Lorentzian reference frame. Averaging over 4-volumes seems more natural than doing so over 3-volumes, but delimiting those is quite problematic (there is no Lorentz-invariant analogy to a Euclidean 3-sphere in 4-D Lorentzian geometry). In general relativistic contexts even the notion of smearing or averaging over a definite 4-volume is in general ill  defined. Therefore, as the problem becomes rather delicate, so does the choice of the quantity that ought to substitute for $r_c$.

An example of the relevance of the preceding discussion and how it might enter the application of CSL to field theoretical contexts, concerns the parameter $r_c$ and how it is employed in the analysis presented in \cite{jmartinPRL}. In Fourier space, the relation between their collapse-generating operator and the parameter $r_c$ is stated to be $\hat{\delta \rho}_\nk = \exp[-k^2 r_c^2/(2a^2) ][ c_1(k) \hat{v}_\nk + c_2(k) \hat{p}_\nk ]$, where $c_i(k)$ are two coefficients, and $\hat v_\nk$, $\hat{p}_\nk$ are quantum and momentum fields respectively. Then, it is argued that because of the presence of the exponential term, the effect of the CSL terms is triggered only once $k$ crosses out the scale $r_c$ , i.e., when its physical wavelength is larger than $r_c$, $k/a < r_c^{-1}$. They obtain from their calculation, depending on the value of $r_c$, that this can happen either during inflation or in the radiation dominated epoch.  However, the statement that \textit{the CSL terms cannot ``localize'' a mode if its ``size'' (its wavelength) is smaller than the localization scale $r_c$} seems quite difficult to accept as a broader argument since it implies that such behavior is to be expected quite generically.

Moreover, we note a transcendent  point: there are conceptual difficulties in attempts to compare two quantities that are of rather different nature: a) the CSL characteristic localization parameter $r_c$, and b) the physical wavelength $2\pi a/k$ associated with a mode $k$.

Let us consider this issue in more detail. In non-relativistic CSL theory, the parameter $r_c$ is a length scale characterizing the level of de-localization associated  with the onset of the spontaneous collapse. For instance, for a single particle, the effect of the collapse aspect of the modified dynamics is negligible if the width of its wave function is much less than $r_{c}$. But, if it is larger than $r_{c}$, the collapse mechanism becomes very relevant.

On the other hand, the wave number of a mode is a byproduct of the Hilbert space characterization of a quantum field and does not characterize, in general, any level of localization. In other words, the CSL theory uses the parameter $r_c$ to represent the length scale for the uncertainty in position of a single particle, where the modification of the dynamics is effectively turned on. The uncertainty in position or the level of spatial de-localization coincides with the width of the wave function $\Psi (x)$. For more general systems, such as quantum field(s), the quantum state might be regarded  as a functional on the space of fields configurations. The connection between the width of the support of such a functional and the spatial localization of the system in physical space are much more complex.

Specifically, when considering a quantum field theory, we should recall that a generic quantum state is characterized, among other things, by quantum expectation values of all the elements of a certain $\mathbb{C}^*$-Algebra of operators \cite{Wald94,Parker,Haag}. In particular, the state would possess some level of uncertainties for general local quantum field operators. Those local operators, including the quantum field itself and its conjugate momenta, are objects which are only distributively defined. In order to consider actual operators in Hilbert space, we need to smear them with smooth test functions of compact support. Furthermore, the formalism must be extended to include the energy-momentum tensor (which involves dealing with the renormalization procedure) and, in fact, go beyond the simple renormalization of its expectation value (the quantity that has been the focus of most studies in that regard), to allow the evaluation and analysis of the corresponding quantum uncertainties.

All of this is highly nontrivial. It is not at all evident that the parameter $r_c$, which in the laboratory setting affects the quantum uncertainties of the position operator, can also be used to characterize, in a simple manner, the relevant uncertainties of the state of the quantum field. Therefore, it is rather unclear how one should extrapolate the use of the parameter $r_c$ to the QFT framework, in such a way that it again serves to encapsulate the ``tolerance of the  theory to de-localized systems'' involving quantum fields, and yet would reduce, in the appropriate limits, to the role it plays in the standard version of CSL. Should that tolerance be associated with each point of the spacetime? That seems unlikely given the distributional nature of the quantum field operators. Perhaps one might consider the use of the $r_c$ parameter, in the context of QFT, as characterizing the size of the smearing region for the operators whose uncertainties the theory drives to reduce. In any case, one can see that the particular set of considerations concerning this issue, employed in the application of the CSL theory to the inflationary universe as done in \cite{jmartinPRL}, is not as natural and direct as one might have thought.

Let us mention a few more issues regarding the landscape of options that one ought to contemplate, when seeking to extend CSL in such a way as to interpolate between present-day laboratory settings and the conditions pertaining to generic quantum fields in curved spacetime appropriate to the inflationary epoch in the early universe. As suggested in e.g. \cite{Diosi1984,diosi1987,diosi1989,penrose1996}, it seems quite natural to think that, at a fundamental level, the spontaneous collapse dynamics might be intimately tied with gravitation. Thus, it would not be at all surprising, if the parameters of the model were effective characterizations of aspects tied to the curvature of spacetime, which resulted from the presence of matter, or more concretely the level of indefiniteness (or uncertainty) of what would correspond to a certain measure of curvature in the fundamental quantum gravity theory.
This in turn, could be naturally expected to be associated with degrees of indefiniteness of the energy-momentum tensor. This would be a natural extrapolation to gravitational theory of the non-relativistic CSL collapse rate dependence on the masses of the particles involved, imposed following its first suggestion in \cite{Pearle1994, Pearle1995}.

There are of course multiple paths to this extrapolation one might consider, but the point is that they are far from unique. One might, for instance, define a local rest frame associated with a state for which the renormalized energy-momentum tensor has limited support in the spatial domain (e.g. as done in the definition of center of mass in GR when that is possible, see \cite{Bblock}) and then perform the smearing over a given volume as defined in that frame. At that point several questions arise: should quantities, like, say curvature, influence the size of the smearing region? Should it affect the value of the collapse rate parameter $\lambda$? As a matter of fact, in a series of works devoted to the examination of the famous black hole information puzzle \cite{BH1,BH4}, it was found that the so called information puzzle might be fully eliminated by the deployment of spontaneous collapse theories. However, for the scheme to work properly one needed to assume a strong dependence of the collapse parameter upon the spacetime curvature. Naturally, there are a large variety of specific quantities on which those parameters could depend, and the exact form of that dependence is, at this point, evidently a fully open issue. Among the possibilities we can list $R$, the Riemann scalar, $R_{ab} R^{ab}$, the square of the Ricci tensor, the Kretchman scalar $R_{abcd}R^{abcd}$, or the magnitude of the Weyl tensor $W_{abcd}W^{abcd}$, etc. If one accepts the validity of the semiclassical Einstein equations, then the first two options would tie the value of the parameters to the renormalized expectation value of the energy-momentum tensor. The last two options will partially decouple them.

And, one should certainly consider that the CSL non-relativistic theory's parameters $\lambda, r_{c}$ might not be \textit{fundamental} constants. Note that the consideration of any of the options mentioned above, in the application of the theory to the cosmological case, implies that dependence on curvature will result in an \textit{effective temporal dependency of the model parameters}. In fact, this kind of effective time dependence (i.e. a time dependence that appears in a certain regime to be a constant, but is in fact a coupling with some other dynamical variable) is one we have already encountered in nature, and specifically in cosmology. For instance, the CSL parameters could possibly depend on quantities characterizing the environment, just as the masses in the standard model of particles depend on a vacuum expectation value in the Higgs sector. Just as the values of relevant fields such as the inflaton and the Higgs fields experienced dramatic changes between the inflationary epoch and the present day, so an effective time dependence of the CSL parameters in the cosmological context might very well be considered likely. Therefore, considering the issue in general terms, one must recognize that there is no particular reason why one should expect that the parameters $\lambda$ and $r_{c}$, utilized in applications of the CSL at present day laboratory situations (and whose values are probably tied to underlying atomic structure that did not exist in inflationary times), should necessarily, or even naturally, be the ones utilized in modeling the inflationary regime as was done in \cite{jmartinPRL}. In our view, a more appropriate approach would be to use the early universe to set constraints on a more general class of plausible versions of CSL type theories involving, as we said, more fundamental objects such as those constructed out of the energy-momentum tensor, introducing natural parameters in those contexts and considering, at the same time, the possible dependence of the parameters on dynamically evolving curvature scalars such as $R$.

In the Appendix, we have constructed a very simple toy model that contains some basic elements of the ideas we have discussed in this section. Namely, the fact that in non-relativistic situations the effective collapse rate seems to depend on the mass of the particle in question might well be, in reality, a special case of a more general dependence involving the space-time curvature. In fact, given that in the context of QFT in curved spacetimes the very notion of particle disappears as a fundamental one \cite{Wald94}, it would seem wholly inappropriate that the mass of a particle would play any fundamental role within that context. In such a profound modification of quantum theory as is involved in spontaneous collapse theories, as long as we work in realms where spacetime notions maintain their viability (i.e. below scales where quantum gravity might be required), it is not unreasonable to assume that spacetime curvature might play a fundamental role. Thus, according to these ideas, it would be clearly inappropriate to compare directly the experimental bounds on $\lambda$, obtained from laboratory situations, with the ones resulting from analyzing the inflationary regime.

Finally, we should mention another problem that appears once one considers the idea of spontaneous collapse theories compatible with \textit{special} relativity\cite{Myrvold}. It is well known that spontaneous collapse theories generically lead to a certain degree of energy production per unit of spacetime volume. In the case of a QFT, one might expect a non-vanishing level of energy creation per mode, but in any truly relativistic version essentially all modes are equivalent (i.e. they are connected by a Lorentz  transformation) so even if we have a finite amount of energy creation per spacetime volume per mode, as the number of modes is infinite, this would translate into an infinite amount of energy creation per spacetime volume; thus, making the theory unviable. One way to remove the problem, as suggested in \cite{Myrvold}, is to make use of exotic degrees of freedom, as is actually done, say, in proposals such as \cite{Bedingham2} or \cite{pearle3}. Another manner to remove the aforementioned problem might be to construct the theory in such a way that the collapse rate depends directly on curvature, so that the rate and thus the energy production would vanish in its absence. In this way, the vacuum state in Minkowski spacetime would not be affected by the collapse at all. The point here is that, even when a single particle is present in the spacetime, the fact that it carries with it some energy-momentum, indicates that strictly speaking (i.e. going beyond the test particle approximation) the spacetime cannot be taken as flat. Consequently, the state of the quantum field is something other than the Minkowski vacuum, and then the collapse dynamics would be ``turned on". The Minkowski vacuum, on the other hand, will be perfectly stable and unaffected by the collapse dynamics.

\section{Conclusions}
\label{conclusions}

In summary, the various points that have been briefly discussed above illustrate the complex issues and several kinds of inconclusiveness one must face
when one considers generalizing the CSL theory from the context of non-relativistic many-particle quantum mechanics, for which it was originally constructed, to the realms of quantum field theory in curved spacetime. Evidently, the latter must be used when one desires an application to the inflationary theory, in particular when focusing on the emergence of seeds of cosmic structure.

Along the way of this exploration, choices must be made. On the one hand, one must choose the setting within which one is going to combine quantum field theory (QFT) with gravitation. While neither approach yet provides a finished program to combine general relativity with QFT, we have argued that, due to the problems of the widely considered approach (i) in the literature, mentioned in Sect. \ref{secdos}, the semiclassical gravity framework (ii) appears favored from a theoretical and conceptual point of view, in particular, when one wants to incorporate collapse models. And, as was also mentioned earlier, predictions using semiclassical gravity are more consistent with the current non-detection of B-modes in the CMB data.

On the other hand, choices must also be made for both the model parameters and the collapse-generating operator. The point we have emphasized throughout this work is that such choices are far from unique, the theory landscape is vast, and in fact we have provided several arguments to cast doubts on some choices sometimes presented as ``the natural ones".

Also, we have analysed and discussed the model presented in \cite{jmartinPRL}. The results obtained by the authors should of course be seen as a consequence of exploring a rather limited and specific approach towards the subject; something briefly addressed in \cite{comment20}. Therefore, one must regard ``the shadow" these authors state their work casts over the general theoretical framework of CSL theory as a rather small one. However, as acknowledged by those authors themselves, without some kind of modification to the standard Quantum Theory, such as that offered by CSL, the account of the emergence of the seeds of cosmic structure during inflation suffers from very serious shortcomings.

The task ahead is to deepen the exploration of the theoretical landscape broadly depicted here, to find a formulation of the theory that reduces to non-relativistic CSL (so far satisfactory for laboratory contexts). Such an extrapolation must be generally covariant, suitable for QFTCS applications and empirically successful in the inflationary context.

\begin{acknowledgements}
G.R.B. is supported by CONICET (Argentina) and he acknowledges support from grant PIP 112-2017-0100220CO of CONICET (Argentina). G.L. is supported by CONICET (Argentina) and the National Agency for the Promotion of Science and Technology (ANPCYT) of Argentina grant PICT-2016-0081. D.S. acknowledges partial financial support from  PAPIIT-UNAM, Mexico (Grant No.IG100120); the Foundational Questions Institute (Grant No. FQXi-MGB-1928); the Fetzer Franklin Fund, a donor advised by the Silicon Valley Community Foundation.

\end{acknowledgements}

\appendix

\section{A toy model where $\lambda$ depends on the spacetime curvature}

As an illustrative example of the ideas considered near the end of section \ref{seccuatro}, let us introduce a toy model in which the CSL parameter $\lambda$ might depend, for instance, on the Kretschmann curvature scalar $K=R_{abcd} R^{abcd}$ (other options could be studied) integrated over a suitable region, which for simplicity we might identify   with the Cauchy hypersurface of collapse. Consider now the case in which we are dealing with a single spinless massive particle in its rest frame. That is, we suppose that
\begin{equation}\label{lambdaRdef}
\lambda(K) = \bigg( \int \sqrt{\gamma}\: d^3 x \: R_{abcd} R^{abcd} \bigg)^{\alpha}
\end{equation}
where $\alpha >0$ and $\sqrt{\gamma}\: d^3 x$ is the 3-volume corresponding to the spatial metric $\gamma_{ij}$, and the integration is over one of the hypersurfaces of constant $t$, i.e. one of the hypersurfaces normal to the static Killing field of the spacetime geometry.

For our toy model, we will assume that the spacetime might be taken the one corresponding to a static spherical object of radius $R$ (\textit{notation warning}: do not confuse this $R$ with the Ricci scalar) with a homogeneous mass density distribution. That is
\begin{equation}
\rho(r)  =
\begin{cases}

\rho_*  & r \leq R \\

0  & r>R

\end{cases}
\end{equation}
Furthermore, assuming an equation of state of the form $p=p(\rho)$, i.e. independent of entropy, it leads to a metric with corresponding line element:
\begin{equation}\label{metricTOV}
ds^2 = -e^{2 \beta(r)} dt^2 + \left[ 1 - \frac{2 G m(r)}{r}   \right]^{-1} dr^2 + r^2 d \Omega^2
\end{equation}
where
\begin{equation}
e^{ \beta(r)} =
\begin{cases}

\frac{3}{2}\left(  1 -\frac{2 G M}{R}  \right)^{1/2}  - \frac{1}{2} \left(  1 -\frac{2 G M r^2}{R^3}  \right)^{1/2}   & r \leq R \\

\left(  1 -\frac{2 G M}{r}  \right)^{1/2} & r>R

\end{cases}
\end{equation}
and
\begin{equation}
m (r)  =
\begin{cases}

\frac{4}{3} \pi r^3 \rho_*   & r \leq R \\

\frac{4}{3} \pi R^3 \rho_* \equiv M  & r>R.

\end{cases}
\end{equation}

Note that the metric characterized by Eq. \eqref{metricTOV} is a consequence of solving the Tolman–Oppenheimer–Volkoff (TOV) equation, which constrains the structure of a spherically symmetric body in static gravitational equilibrium, within GR.

Using \eqref{metricTOV} we calculate $K \equiv R_{\alpha \beta \gamma \delta} R^{ \alpha \beta \gamma \delta }$ in two regimes: For $r \leq R$
\begin{eqnarray}\label{Kmenor}
& & K_<   = \frac{12 r_s^2  }{R^6 \left(8 R^3-9 R^2 r_s+r_s r^2\right)^2} \nn
&\times& \bigg[-153 R^5 r_s+81 R^4 r_s^2-9 R^2 r_s^2 r^2 +2 r_s^2 r^4 \nn
&+& R^3 r_s r^2 \left(5-6 \sqrt{1-\frac{r_s}{R}} \sqrt{1-\frac{r_s r^2}{R^3}}\right)\nn
&+&R^6 \left(6 \sqrt{1-\frac{r_s}{R}} \sqrt{1-\frac{r_s r^2}{R^3}}+74\right) \bigg]
\end{eqnarray}
and for $r > R$
\begin{equation}\label{Kmayor}
K_>  = 12 \frac{r_s^2 }{r^6},
\end{equation}
where $r_s \equiv 2 G M $.

With the Kretschmann scalar at hand, and taking into account the volume element $\sqrt{\gamma} d^3x$, we can compute $\lambda(K)$. That results in
\begin{eqnarray}\label{inttotal}
\lambda(K)^{1/\alpha} &=& 4\pi \int_0^R  dr \: K_< \left[ 1 - \frac{r_s r^2}{R^3}   \right]^{-1/2}  r^2 \nn
& +& 4\pi \int_R^\infty dr\: K_> \left[ 1 - \frac{r_s}{r}   \right]^{-1/2}  r^2 \nn
&\equiv& \mathcal{I}_1 + \mathcal{I}_2.
\end{eqnarray}
Both integrals, $\mathcal{I}_1$ and $\mathcal{I}_2$, can be calculated analytically.

With $M$  the mass of a particle, and  taking $ R \gg r_s$ we perform a Taylor expansion of $\mathcal{I}_1$ and $\mathcal{I}_2$. The resulting dominant term is given by
\begin{equation}\label{I1app}
\mathcal{I}_1  + \mathcal{I}_2   \approx  36 \pi \frac{r_s^2}{ R^3}  = \frac{144 \pi G^2 M^2}{R^3}.
\end{equation}

 If one assumes that $R$ is to be identified with some fixed length scale, which might be related to the $ r_c$ parameter of the CSL theory, then for $\alpha = 1/2$ one finds $\lambda(K) \simeq M$. One obtains a similar result (as far as the $M$ dependence is concerned) if, instead of the Kretschmann scalar, one takes $W_{abcd}W^{abcd}$, where $W_{abcd}$ stands for the Weyl tensor. Adopting such a proposal might be natural if framed within what would seem as rather dramatic reconsideration of how single elementary particles gravitates (i.e. not as a point source but as an extended one), a point of view that might not be as problematic when considering that the situation we would be envisioning (when collapse becomes important) involves effective particle's wave functions with position uncertainties of the order of  $r_c$ or even larger.

Another possibility that might seem rather natural involves identifying the ``radius'' $R$ with the particle's Compton wavelength, i.e. $R \sim 1/M$. In that case  we find
\begin{equation}\label{I2app}
\mathcal{I}_1  + \mathcal{I}_2   \approx  36 \pi \frac{r_s^2}{ R^3}  \sim 144 \pi G^2 M^5
\end{equation}

In this case, setting $\alpha = 1/5$ one again obtains $\lambda(K) \simeq M$. Once more, a similar result is obtained as far as the $M$ dependence is concerned, if instead of the Kretschmann scalar one uses $W_{abcd}W^{abcd}$.

These examples illustrate aspects of some paths that might be taken in constructing a general version of CSL applicable to QFT in curved spacetimes, and from which the effective mass proportionality of the usual non-relativistic version of CSL (developed with laboratory conditions in mind) could possibly arise. Needless is to say that   the general issue is, of course, far from being fully studied by the simple examples considered in this Appendix.

\bibliography{bibliografia}
\bibliographystyle{apsrev}

\end{document}